\documentclass[aps,prb,twocolumn,reprint,showpacs,superscriptaddress,amsmath,floatfix]{revtex4-1}
\usepackage{graphicx, epsfig, ulem, dcolumn, bm}
\begin{document}

\title{Reentrant cluster glass and stability of ferromagnetism in Ga$_2$MnCo Heusler alloy}
\author{Tamalika Samanta}
\affiliation{Department of Metallurgy Engineering and Materials Science, Indian Institute of Technology Indore, Simrol, Indore 453 552, India}
 \author{P. A. Bhobe}
 \email{pbhobe@iiti.ac.in}
\affiliation{Department of Metallurgy Engineering and Materials Science, Indian Institute of Technology Indore, Simrol, Indore 453 552, India}
\affiliation{Discipline of Physics, Indian Institute of Technology Indore, Simrol, Indore 453 552, India}
\author{A. Das}
\affiliation{Solid State Physics Division, Bhabha Atomic Research Centre, Mumbai 400 085, India}
\author{A. Kumar}
\affiliation{Solid State Physics Division, Bhabha Atomic Research Centre, Mumbai 400 085, India}
\author{A. K. Nigam}
\affiliation{Tata Institute of Fundamental Research, Homi Bhabha Road, Mumbai 400 005, India}

\date{\today}

\begin{abstract}
We present here a detailed investigation into the magnetic ordering of full Heusler alloy Ga$_2$MnCo using dc, ac magnetization measurements,  neutron diffraction and neutron depolarization experiments. Crystal structure at room temperature was first confirmed to be L2$_1$ using the highly intense synchrotron X-ray diffraction (XRD) technique. Temperature dependent magnetization reveals that Ga$_2$MnCo enters a ferromagnetic (FM) state at $T_C=$154 K, characterized by a sharp increase in magnetization and a plateau-like region hereafter. As the temperature is decreased further, a sharp drop in magnetization is observed at $T_f$ = 50 K, hinting towards an antiferromagnetic (AFM) phase change. Neutron diffraction (ND) recorded over the range of temperature from 6 to 300 K, provides combined information regarding crystal as well as magnetic structure. Accordingly, an increase in the intensity of the ND pattern is seen at 150 K, signaling onset of long range FM order. However, there is no sign of appearance of superlattice reflections corresponding to the AFM phase, in the patterns recorded below 50 K. An unusual discontinuity in the unit cell volume is seen around $T_f$ indicating a coupling of this second transition with the contraction of the lattice. Attempts to unravel this interesting magnetic behaviour using ac susceptibility measurements lead to the existence of glassy magnetism below $T_f$. Systematic analysis of the susceptibility results along with neutron depolarization measurement, identifies the low temperature phase as a reentrant cluster glass.
\end{abstract}

\maketitle

\section{\label{sec:level1}Introduction}
Generally, full Heusler alloys are of the chemical form X$_2$YZ, where X and Y are transition metals and Z is a main group element belonging to the $s$ and $p$ blocks. The crystallographic positions available to these atoms are A($\frac{3}{4}$, $\frac{3}{4}$, $\frac{3}{4}$); B($\frac{1}{2}$, $\frac{1}{2}$, $\frac{1}{2}$); C($\frac{1}{4}$, $\frac{1}{4}$, $\frac{1}{4}$); D(0,0,0). Accordingly, L2$_1$ structure is realized when X atom occupies A and C sites, Y atom occupies the B site and Z atom is present at D site resulting in XYXZ order. Presence of two magnetic sublattices in a full Heusler system often results in magnetic interactions such as anti-ferro, ferri, compensated ferrimagnetism to localized-itinerant ferromagnetism including complex spin glass state \cite{kubler, mei, rolf, zhang}. Hence apart from being technologically useful, they are ideal systems to study the nature of magnetic correlations in diverse magnetically ordered states. Unlike such standard Heusler form, the Ga$_2$MnCo composition has a $sp$ element in excess with 2 parts of Ga to 1 part each of Mn and Co, reversing the general chemical formula to Z$_2$XY, yet maintaining the L2$_1$ structural form.

Very few examples of such Z$_2$ based systems exist which includes the ferromagnetic shape memory alloy Ga$_2$MnNi with a martensitic transformation at 780 K and a Curie temperature at 330 K \cite{barman} and Al$_2$MnCo\cite{yamaguchi} with a re-entrant spin glass behaviour and a dominating ferromagnetic coupling between transition metals. First principle electronic structure calculations have also been performed on isostructural alloying of Ga$_2$MnCo -- Ga$_2$MnV alloys, where the exchange interaction between Mn and V atoms is found to change from antiferromagnetic to ferromagnetic coupling with increasing c/a \cite{chen}. This gives an important indication of finding myriads of unusual magnetic and structural properties on exploring new Z$_2$-based Heusler alloys.

Magnetism in Ga$_2$MnCo is considered to be of ferrimagnetic type, originating from the antiparallel aligned moments of Mn and Co sublattices\cite{chen, li}. The other two Heusler compositions in the Co-Mn-Ga alloys, \textit{viz}. Co$_2$MnGa (L2$_1$) and Mn$_2$CoGa (Hg$_2$CuTi) show strong ferromagnetic and soft ferrimagnetic behaviour respectively\cite{hames, ali}.  Infact, in Ref. \cite{li}, a magnetic compensation behaviour was found with systematic substitution of Ga in Co$_2$MnGa as the chemical composition changed from Co$_2$MnGa to Ga$_2$MnCo and was conjentured to be due to competitive nature of the Co and Mn moments. Furthermore, band structure studies reveal a stronger covalent hybridization between the $p$-electrons of main group Ga atoms and $d$-electrons of transition-metal atoms. This hybridization is believed to cause a higher density of states (DOS) near Fermi level (\textit{E$_F$}), and subsequently lead to an unstable magnetic configuration \cite{li}.

Early work on GaCo$_{1-x}$Mn$_x$ ($x=0-0.55$) pseudo-binary compositions \cite{shiriasi} claimed it to be a reentrant spin glass based on ac magnetic susceptibility measurements carried out under dc magnetic fields. It may be noted that the investigated compositions were not in 2:1:1 Heusler ratio. Atomic disorder does lead to a \textit{glassy} magnetic state in many systems \cite{mydosh} and can coexist with exotic ground state such as superconductivity, as found in pseudobinary intermetallic compounds \cite{huser}. Other example includes the giant exchange bias field with the reentrant spin glass - FM interface caused by anti-site disorder in Mn atoms in Mn$_2$Ni$_{1.6}$Sn$_{0.4}$ Heusler alloy\cite{ma}. Such glassy behaviour (mostly non-canonical) with exceptional functional properties leads to a resurgence of attention in this area. Thus it appears that the problem of magnetic order in Ga$_2$MnCo has a nontrivial interest in its own right since there exists excess main group atoms besides two magnetic atoms, that play a significant role in the formation of magnetic ground state.

With an aim to obtain a deeper perspective about magnetic correlations in Ga$_2$MnCo, we have carried out a thorough investigation into its crystal structure, magnetic, and electrical transport properties. A combination of high-resolution synchrotron X-ray diffraction and neutron diffraction study clarifies the crystal structure and exact site occupancies of the magnetic ions in Ga$_2$MnCo. DC magnetization and the neutron depolarization measurements carried out as a function of temperature, identify the distinctive magnetic phase transition temperatures. Temperature variation of ac susceptibility unravel the low temperature glassy phase. In this paper, we discuss the role of magnetic ions in pushing the system towards a glassy magnetic state at low temperatures even after ordering as a ferromagnet at higher temperatures.

\section{\label{sec:level2}Experimental Details}
Polycrystalline bead of Ga$_2$MnCo (6.5 g) was prepared by arc-melting the starting elements ($\ge$99.99\% purity) under argon atmosphere. To ensure good homogeneity, the bead obtained after first melting was flipped over and remelted several times ensuring minimum weight loss. The homogeneous bead so obtained was sealed in evacuated quartz tube and annealed at 800$^{\circ}$C for 7 days before quenching in ice water. The phase purity of the sample was checked by powder diffraction recorded at room temperature using synchrotron radiation ($\lambda$ = 0.9782 \AA) at Indian beamline, BL18B, Photon Factory, KEK, Japan. Energy dispersive X-ray analysis (EDAX) using SUPRA 55 Zeiss field emission scanning electron microscope confirms the homogeneity of the sample. The elemental ratio 51.4 : 25 : 23.7 for Ga : Mn : Co, represents an average value (with a standard deviation of 3\%) obtained after recording several EDX spectra from different spatial locations on the sample. The four probe electrical resistivity measurements were carried out using the transport properties option of Quantum Design Inc. physical properties measurement system. The magnetization measurement was done using a SQUID-based magnetometer, by Quantum Design Inc. The ac susceptibility as a function of temperature was measured at various frequencies using a physical properties measurement system. Neutron diffraction patterns were recorded on the PD2 powder neutron diffractometer ($\lambda$=1.2443 \AA) at the Dhruva reactor, Bhabha Atomic Research Centre, Mumbai. Neutron depolarization measurements were carried out on the Polarized Neutron Spectometer at Dhruva reactor.

\section{\label{sec:level3}Results}
\begin{figure}
\begin{center}
\includegraphics[width=1\columnwidth]{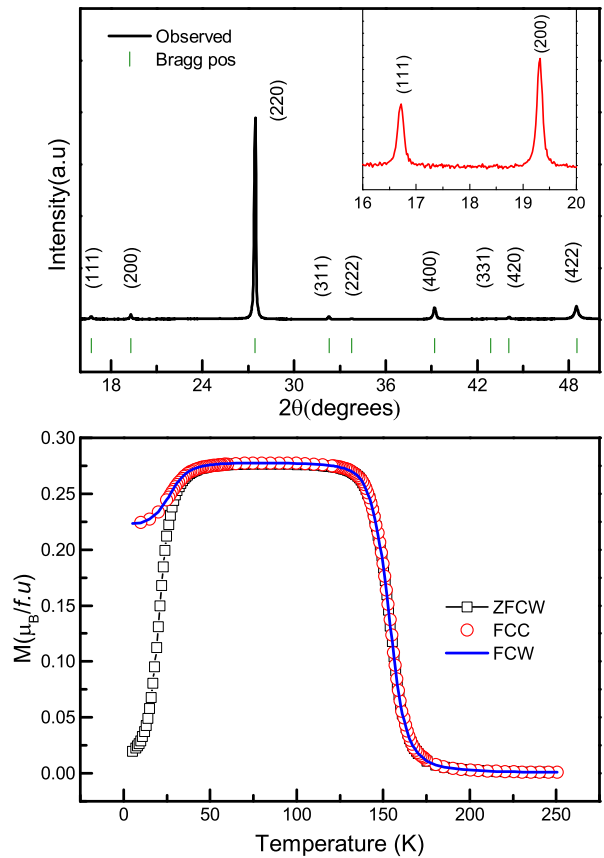}
\end{center}
\caption{\label{xrd-MT}(Color online) Structural and Magnetic properties of Ga$_2$MnCo. (a) The room temperature synchrotron powder XRD pattern of Ga$_2$MnCo. The bars show the Bragg positions. Inset: Magnified view of the (111) and (200) superlattice reflections. (b)Temperature dependence of magnetization of Ga$_2$MnCo in an applied magnetic field of 100 Oe.}
\end{figure}
Ga$_2$MnCo was examined for its crystal structure using the high-resolution synchrotron X-ray powder diffraction (XRD) with wavelength $\lambda$ =0.9782 \AA. The XRD profile recorded at room temperature is presented in the Fig. \ref{xrd-MT}(a). All the observed peaks are indexed to a face centered cubic cell in \textit{$Fm\bar{3}m$} space group, confirming the phase purity of the sample. Besides the principle reflection peaks of (220), (400), (422) planes, the fcc-typical superlattice reflections of (111) and (200) planes are also present (see inset), indicating Ga$_2$MnCo to be a well ordered, single phase system with cubic L2$_1$ structure. However, the intensity of these reflections is adversely affected due to similar X-ray scattering factors for Mn and Co atoms. The L2$_1$ structure is described as four interpenetrating \textit{fcc} sublattices with X atoms occupying the \textit{8c} wyckoff positions (0.25, 0.25, 0.25) and, Y, Z atoms located in \textit{4b} site (0.5, 0.5, 0.5) and \textit{4a} site (0, 0, 0), respectively. Hence, the presence of (111) and (200) peaks signifies a highly ordered L2$_1$ lattice. Further, from the XRD pattern we estimate the lattice parameter of the sample to be 5.828 \AA. To circumvent the problem of low intensity of the superlattice reflections in the XRD pattern, we undertake a thorough structural investigation using neutron diffraction, as discussed later in the text.

\begin{figure}
\begin{center}
\includegraphics[width=1\columnwidth]{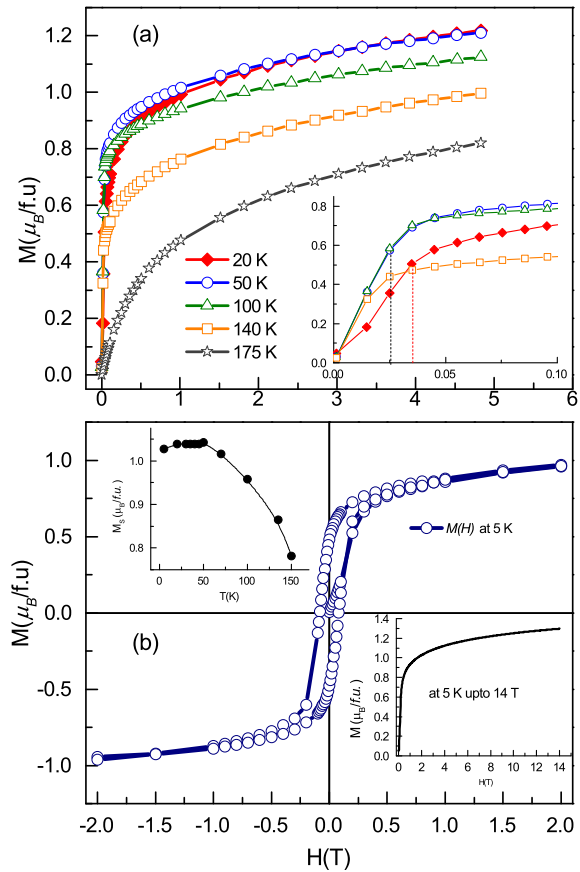}
\caption{\label{MH-1}(Color online) Isothermal magnetization for Ga$_2$MnCo recorded at different temperatures. (a) \textit{M(H)} curves upto a field of 5 Tesla at 20 K to 175 K. Inset: Magnified view of the \textit{M(H)} curves (upto 0.1 T) (b) Magnetic hysteresis loop of Ga$_2$MnCo at 5 K, and H = 2 T. Left inset: Variation of magnetic saturation moment recorded at different temperatures.
Right inset: \textit{M(H)} curve for Ga$_2$MnCo at 5 K and H upto 14 T.}
\end{center}
\end{figure}

Next, we study the magnetic properties of Ga$_2$MnCo through DC magnetization measurement carried out as a function of temperature and applied magnetic field. Magnetization as a function of temperature (\textit{M(T})) recorded in an applied field of 100 Oe, following the ZFCW (zero-field-cooled-warming), FCC (field-cooled-cooling) and FCW (field-cooled-warming) protocols, is shown in Fig. \ref{xrd-MT}(b).  A signature of ferromagnetic ordering is seen at \textit{T$_C$}  = 154 K, followed by an abrupt downturn in \textit{M(T)} below 50 K, hinting a crossover to an antiferromagnetic order. We label this second transition as \textit{T$_s$}. A clear bifurcation in ZFCW - FC curve occurs in this low temperature region indicating non-ergodicity in the M(T) behaviour of Ga$_2$MnCo below \textit{T$_s$}. Figure \ref{MH-1}(a) shows the dc magnetization measurement with field variation (\textit{M(H)}), recorded at different temperatures spanning the \textit{T$_C$} and \textit{T$_s$}. All the \textit{M(H)} plots recorded at temperatures below \textit{T$_C$} are quick to attain near-saturation values at a relatively smaller applied field of $\sim$ 0.1 T.  Comparatively, curves recorded at temperatures below \textit{T$_s$} show a sluggish built up towards saturation (see inset of Fig. \ref{MH-1}(a)). Overall, the saturation values, $M_s$, estimated by extrapolating the \textit{M(H)} to H = 0, seem to increase systematically below \textit{T$_C$} and attain a stable value below \textit{T$_s$}. This observation is represented in the left inset of Fig. \ref{MH-1}(b).

Importantly, as can be seen in Fig. \ref{MH-1}(b), the \textit{M(H)} at 5K shows considerable hysteresis like a ferromagnet but does not saturate even under an applied field of 14 T, shown in the right inset of \ref{MH-1}(b). Due to the presence of AFM interactions, a portion of the atomic moments is aligned anti-parallel to the direction of spontaneous magnetization. Rotation of these moments in large magnetic fields determines the linear dependence of magnetization above saturation. The total saturation moment at 5 K, estimated by extrapolating this \textit{M(H)} to H = 0, is 1.12$\mu_B$/f.u. and matches with the reported value of 1.34 $\mu_B$ recorded by Ref. \cite{li, chen}.
However, this observed value differs considerably from the value calculated using the Slater Pauling rule \cite{gala}. Here, the saturation magnetization is estimated using a simple relation, M$_S$ = [$N-24$], where $N$ represents total number of valence electrons present in the system. Following this expression, the saturation magnetic moments obtained for stoichiometric Co-Mn-Ga alloys \textit{viz.} Co$_2$MnGa, Mn$_2$CoGa, turn out to be 4$\mu_B$, 2$\mu_B$, and match with their experimentally reported \cite{li} values of 4.14$\mu_B$, 2.1$\mu_B$, respectively. While, experimental value for Ga$_2$MnCo seems to deviate from its estimate of 2$\mu_B$, hinting a re-look into magnetic interactions in Ga$_2$MnCo.

\begin{figure}
\begin{center}
\includegraphics[width=1\columnwidth]{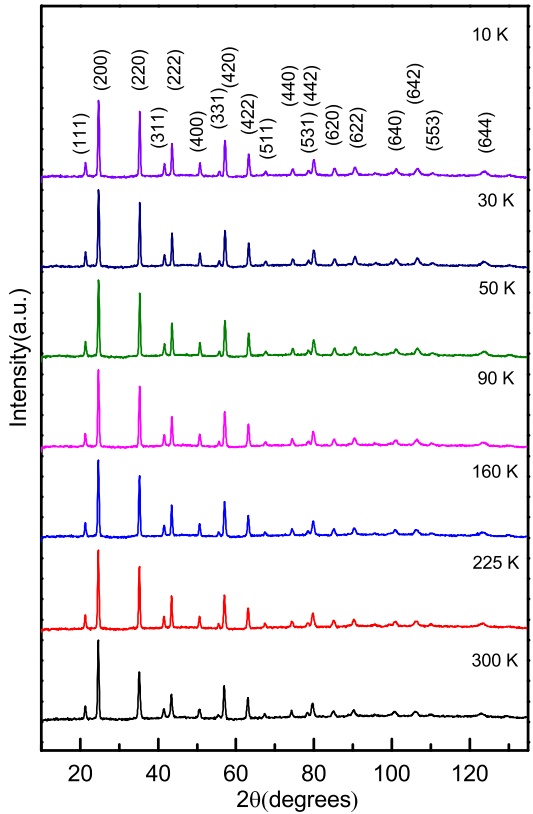}
\caption{\label{neutron}(Color online)ND pattern of Ga$_2$MnCo at various temperatures.}
\end{center}
\end{figure}
\begin{figure}
\begin{center}
\includegraphics[width=1\columnwidth]{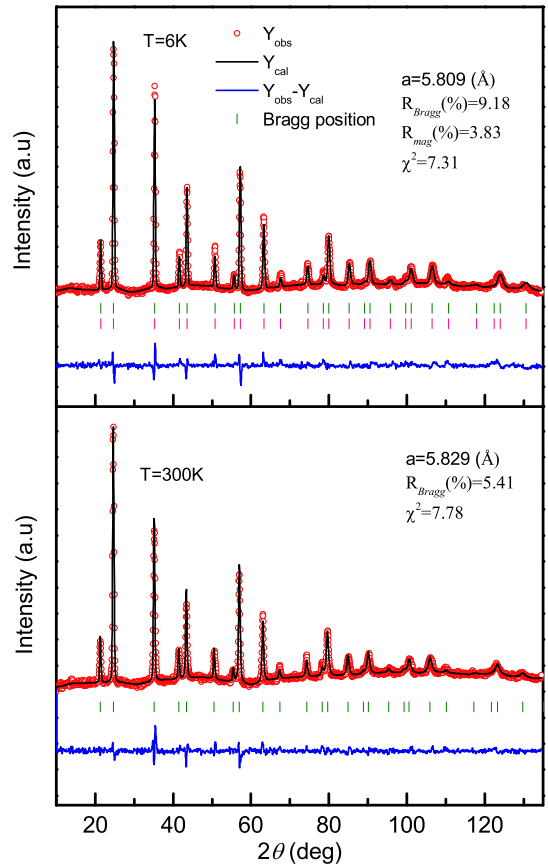}
\caption{\label{neutron-1}(Color online)The Rietveld refinement of ND pattern of Ga$_2$MnCo at 6 K and 300 K. Experimental data are represented by open circles, the calculated profile by a continuous line, and the allowed structural and magnetic Bragg reflections by vertical marks. The difference between the experimental and calculated profiles is displayed at the bottom of the graph.}
\end{center}
\end{figure}

Complex magnetic interactions below \textit{T$_s$}, different from the one in the region $T_s \le$ T $\le T_C$ might exist in Ga$_2$MnCo, which require further explorations. With this aim, we undertook powder neutron diffraction study of Ga$_2$MnCo at various temperatures (6 K $\leq$ T $\leq$ 300 K) spanning all its magnetic transitions. The raw diffraction patterns are presented at Fig. \ref{neutron}. Firstly, no extra peaks are observed in the diffraction patterns recorded below \textit{T$_s$}, ruling out any long range AFM order. Also, no signature of any structural distortions is evident from any of the patterns. The quantitative analysis carried out using Reitveld refinement method and implemented through FULLPROF suite \cite{fullprof}, reaffirms the $Fm\bar{3}m$ symmetry in the crystal structure. Representative fits for diffraction patterns at 6 K and 300 K are shown in Fig. \ref{neutron-1}. The unit cell parameter at room temperature is found to be 5.829 \AA,~ in agreement with the XRD data.

Unlike X-rays, scattering with neutrons allows much better differentiation of site occupancy of the constituent atoms since the nuclear coherent scattering amplitudes of Co, Mn are widely different. We observe that a good match between the recorded data and the calculated profile is obtained albeit after inclusion of $\sim$ 10 \% of site swapping between Mn and Co Wyckoff positions. Refinement of the data using site-disorder of Ga atoms with Mn/Co was also explored, but yields a poor fit to the experimental data. As the diffraction patterns are analyzed with respect to changing temperature, the intensity of low angle reflections show appreciable increase below \textit{T$_C$}  = 154 K, due to the FM long-range order. Below $T_C$, the refinement of the powder diffraction data has been carried out taking into account both the nuclear and magnetic phases. The parameters varied are scale, cell, background, moments on Co and Mn and overall thermal parameters. The magnetic structure is found to be FM below $T_C$ with magnetic moment of 0.8(0.1) $\mu_B$ and 0.6(0.1) $\mu_B$ on Mn and Co, respectively at 6 K. However, the temperature variation of the moment obtained from ND do not show the drop in magnetization below 50 K in \textit{M(T)}. Also, no superlattice reflection corresponding to a long-range AFM order appears below 50 K. Therefore combining magnetization and neutron diffraction studies we conclude that the long-range FM order that develop below \textit{T$_C$} is hampered due to development of some short-ranged AFM correlations below 50 K, giving rise to a situation like that of a cluster glass phase.

In order to obtain a rough estimate of the size of such FM clusters, we carried out neutron depolarization experiment on the same powdered sample. Polarization is simply related to the Larmor precision of the neutron spin around the magnetic induction $B$ in the sample. In the paramagnetic state the spin fluctuates over a very short-time scale compared to the typical Larmor time for the precession and the neutron spins do not follow the $B(t)$ variation. Therefore, no depolarization is observed. Similarly, a spin glass state with atomic level of magnetic inhomogeneity also does not depolarize the neutron beam as $B(t)$ averages out to zero on spatial scale. Whereas, the incident neutron gets depolarised by traversing through randomly oriented domains as in a typical ferromagnet. This technique for studying FM domains was initiated by Halperin and Holstein \cite{hal} and since been able to give useful information on the magnetic inhomogeneity with mesoscopic length scale like spin clusters carrying net moment.

\begin{figure}
\begin{center}
\includegraphics[width=1\columnwidth]{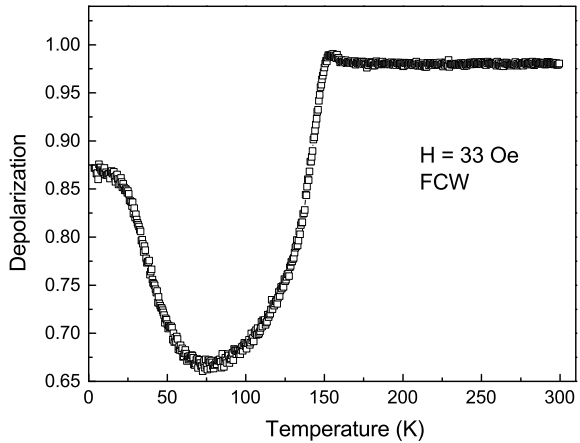}
\caption{\label{NDNEW}(Color online) Temperature dependence of the transmitted neutron polarization in H=33 Oe.}
\end{center}
\end{figure}

Temperature dependence of neutron beam polarization in Ga$_2$MnCo was recorded across its paramagnetic, FM and the low temperature phase. The measurement was performed in FCW scan where the sample was cooled under the external field of 33 Oe and the data was recorded while warming. As can be seen from Fig. \ref{NDNEW}, with the decrease in temperature the polarization starts decreasing sharply below 160 K, signifying the onset of long range FM order. Thereafter, a plateau-like region appears in the region between $\sim$ 128 K to $\sim$ 50 K. This entire trend in polarization resembles a mirror-image of the $M(T)$ measurement recorded in FC mode. Below 50 K the polarization again increases, denoting either a weakening of domain size or domain magnetization. It may be noted that the polarization does not recover to its paramagnetic value, hence long-range ferromagnetic order is not lost. As seen from the inset to Fig. \ref{MH-1}(b), the spontaneous magnetization does not decrease below $\sim$ 50 K. Therefore, the increase in polarization below 50 K could be due to weakening of the domain size.

An estimate of the domain size in the ferromagnetically ordered region of temperature can be obtained  from the expression,
 $$P_f = P_i exp\Big[- \alpha(\frac{d}{\Delta})\langle\Phi_{\delta}\rangle^2\Big]$$
where $P_i$ and $P_f$ are initial and final beam polarization, $\alpha$ is a dimensionless parameter set to 1/3, $d$ is the sample thickness, $\Delta$ is the typical domain size and $\Phi_{\delta}$ = (4.63 $\times$ 10$^{-10}$ G$^{-1}$\AA$^{-2}$)$\lambda B\Delta$ is the precession angle and $\lambda$ is the neutron wavelength. The bulk magnetization values presented at Fig. \ref{MH-1} is used here to obtain the domain magnetization $B$. Using the above expression we obtain an average domain size of $\sim$ 2$\mu$m at 75 K. However, it may be noted that the value quoted here is a fairly close estimate, and evaluation of more accurate value would require precise determination of $B$ and three dimensional polarization analysis \cite{das-depol}. Nevertheless, it is seen that large FM domains exists in Ga$_2$MnCo indicative of long-range magnetic order and transition below 50 K shortens the domain size. To sum up, the neutron diffraction and magnetization measurement indicate that the long-range FM interactions that develop below 154 K are interrupted and their spatial extent shortened by some AFM interactions that develop below 50K. Thus large FM clusters are connected via short-range oppositely aligned magnetic linkages.

\begin{figure}
\begin{center}
\includegraphics[width=1\columnwidth]{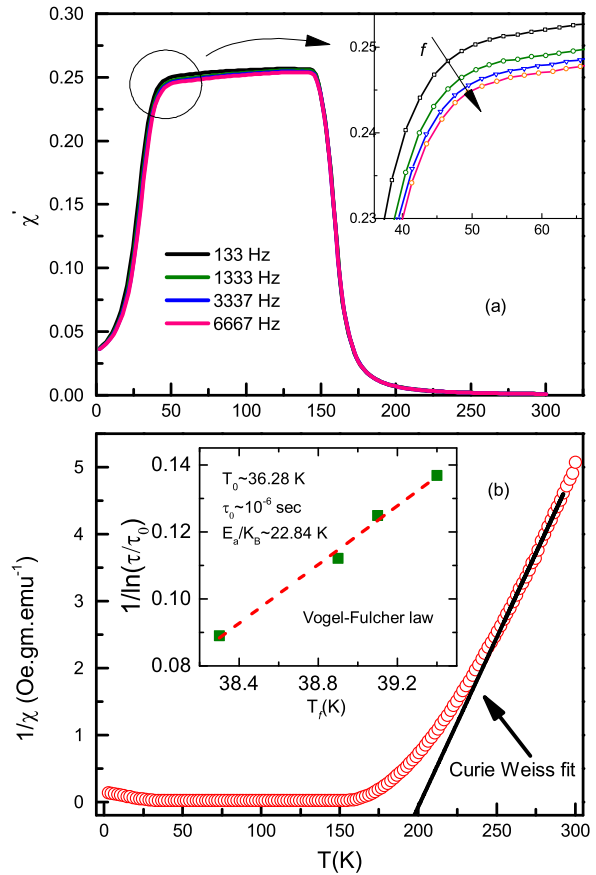}
\caption{\label{chi-1}(Color online) Upper panel shows the temperature dependence of the real part of the ac susceptibility measured for Ga$_2$MnCo between 2 -- 300 K, at different frequencies. The inset shows the magnified view of the $chi$ at T$_f$. Lower panel shows Curie-Weiss fit to the susceptibility of Ga$_2$MnCo. Inset: The variation of the freezing temperature T$_f$ with the frequency of the ac field in a Vogel-Fulcher plot with $\tau=10^{-6}$. Dashed line is the fit to the Vogel-Fulcher equation.}
\end{center}
\end{figure}
To gain better insight into the origin of the second magnetic transition seen at $T_s$, low field ac susceptibility was measured. The temperature dependence of $\chi_{ac}$ was recorded at different frequencies ($f$) ranging from 133 Hz to 6667 Hz. Figure \ref{chi-1}(a) shows the real component of the ac susceptibility ($\chi_{ac}^{\prime}$), where the overall trends and transition temperatures match with the previously stated dc $M(T)$ measurement. What is important to note is that the feature associated with $T_s$ shifts towards higher temperature as the frequency changes from 133 Hz to 6667 Hz. We observe a change in temperature values from 38.3 K at 133 Hz to 39.5 K at 6667 Hz. The frequency dependence of \textit{T$_s$}, though weaker, is similar to that of a Ruderman-Kittel-Kasuya-Yosida (RKKY) spin system \cite{tholence}. In the lower panel of the Fig. \ref{chi-1} a linear fit to the inverse susceptibility curves yields $\theta_{CW}$=198 K, a value much greater than T$_C$ (= 154 K),  which suggests that the spins begin to align locally much before a long range FM order is realized. This observation also clarifies the curvature in the \textit{M(H)} recorded at 175 K (see Fig. \ref{MH-1}(a)) in contrast to a typical paramagnetic state.

The frequency dependent shift in T$_s$ gives a clear indication of a glassy magnetic phase being present at low temperature in Ga$_2$MnCo. Rather, to be more apt, $T_s$ signifies the onset of a re-entrant spin glass (RSG) in Ga$_2$MnCo as the glassy state emerges after the system has ordered in a stable FM state \cite{ma, motoya}. The flat topped hump like feature at $T_s$, seen in the $\chi_{ac}^{\prime}$ vs. T curve is suggestive of the critical concentration of magnetic entities for percolation of ferromagnetism \cite{cole}. A frequency dependent shift in the temperature associated with this feature is indicative of formation of cluster-like entities of varying sizes, and the temperature itself can be associated with the spin freezing temperature, $T_f$. The spin freezing/blocking process is defined by $\phi$ = $\Delta$T$_f$/[T$_f$ $\Delta$ log$_{10}$(\textit{f})]  and  $\phi$ varies from approximately 0.005 to 0.05 depending on the systems \cite{mydosh}. For a well known canonical spin glass such as CuMn,  $\phi$  is estimated to be 0.005. For Ga$_2$MnCo $\phi$ works out to be 0.017, indicating a much larger sensitivity to the frequency and qualifying it to be classified in the cluster glass regime \cite{myd}.

To further verify the cluster glass state in Ga$_2$MnCo, a logarithmic frequency dependence of the freezing temperature that follows Vogel-Fulcher empirical law, is proposed. As per this law, $\tau = \tau_0 exp[\frac{E_a}{k_B(T_f - T_0)}]$, where $\tau_0$ is the characteristic time, $E_a$ and $T_0$ are the activation energy and Vogel - Fulcher temperature, respectively,  that gives inter-cluster interaction strength. Inset to Fig. \ref{chi-1}(b) shows the Vogel-Fulcher plot for Ga$_2$MnCo along with the extracted values for $E_a$/K$_B$, $T_0$ and characteristic time $\tau_0$. The data for Ga$_2$MnCo gives a $\tau_0$ value of 10$^{-6}$ sec, which is much higher than that obtained for conventional spin glasses ($\sim$ 10$^{-12}$ to 10$^{-10}$ sec) \cite{mydosh}. Such a value of characteristic time which is also found in RSG systems like Ni$_2$Mn$_{1.36}$Sn$_{0.64}$ \cite{sma} suggests a slow spin dynamics in Ga$_2$MnCo due to the cluster formation. The values of $E_a/K_B$ and $T_0$ are found to be 22.84 K and 36.28 K respectively. $T_0$ is very close to the value of spin freezing temperature obtained from linear ac susceptibility measurements indicating that the RKKY interaction is relatively strong in this compound. Tholence criterion \cite{tholence} $t^* = \frac{(T_f - T_0)}{T_f}$ for Ga$_2$MnCo works out to be 0.052. Altogether these observations clearly suggest that the spin glass state in Ga$_2$MnCo is related to the FM cluster formation and falls in a category of a Ruderman-Kittel-Kasuya-Yosida (RKKY) spin glass system.

\section{\label{sec:level4}Discussions}
\begin{figure}
\begin{center}
\includegraphics[width=1\columnwidth]{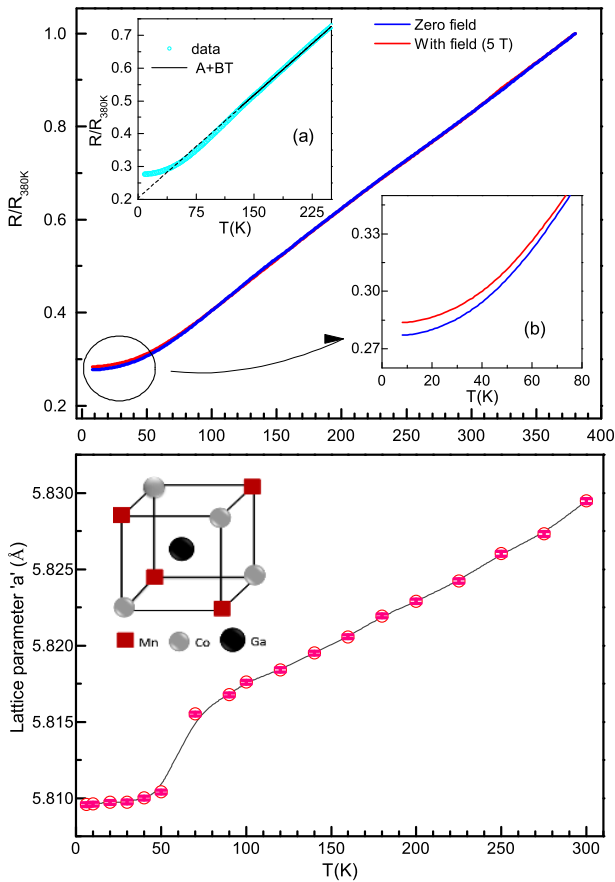}
\caption{\label{Res}(Color online) Upper panel: Dependence of normalized resistance as a function of temperature, with and without the presence of magnetic field. The left inset (a) shows the phonon contribution. The right inset (b) shows the magnified view of the resistance plot at low temperatures. Lower panel: The lattice parameter variation with temperature as obtained from refining the ND data. Inset shows the L1$_0$ configuration containing the magnetic ions. }
\end{center}
\end{figure}

Neutron diffraction revealed that Ga$_2$MnCo lattice maintains the cubic structure over the entire temperature range and the variation of lattice parameter with temperature is plotted in the lower panel of Fig. \ref{Res}. These values are found to vary from 5.8095(3) \AA~ at 6 K to 5.8294(4) \AA~ at 300 K. Surprisingly, an abrupt change of about 0.1\% is found in the lattice parameter around the cluster glass phase transition, 50 K. Beyond 50 K, the lattice undergoes normal thermal expansion with increasing temperature.

Although the degree of lattice expansion at $T_s$ is small, we tried to find its evidence, if any, in the transport properties of Ga$_2$MnCo. For this, the electrical resistance was measured by varying the temperature of the sample, as depicted in Fig. \ref{Res}. Conventional metallic conduction behaviour i.e. $\frac{dR}{dt} > $ 0 is observed throughout the measured temperature range. The data seems to follow a typical phonon dependence ($\propto$ T) for most part of the temperature: 380 to 130 K. Below 130 K, ($<$ $T_C$) the slope of the curve starts to change slightly as the electron electron interaction and magnetic correlation effects become significant. No sign of any anomaly related to the lattice distortion at $T_s$ is observed. Application of sufficiently high magnetic field of 5 T leads to about 2.5 \% increase in the resistance below $T_s$ (see inset to Fig. \ref{Res}).

Though there is no symmetry breaking transition witnessed from neutron diffraction measurement, the cell volume expansion at the cluster glass transition temperature indicates correlation of magnetic and crystal structural degrees of freedom. The origin of such anomaly should lie in the $\sim$ 10\% disorder in the site occupancy of Co and Mn crystallographic positions, encountered during refinement of neutron diffraction profiles. We believe the presence of this site disorder decides the extremity of the magnetic interactions which leads to the associated cluster glass state. Inset of Fig. \ref{Res} (lower panel) presents a schematic of the L1$_0$ sub-cell of magnetic atoms in the Ga$_2$MnCo that forms the part of larger L2$_1$ unit cell. In the ideal lattice scenario, Mn atoms have Co as their nearest neighbour(NN) magnetic atoms and other Mn atoms present along the face diagonal as their next nearest neighbours(NNN). This NNN Mn--Mn interaction is FM in nature that orders below 154 K, as witnessed from the magnetization and neutron diffraction measurement. The non-magnetic Ga atom sits at the body centered position in the sub-cell.

Our results suggest that, in the event where one of these NN Co atoms swaps position with Mn, we have a situation with a new NN correlation identified as Mn--Mn$'$. The swapped Mn$'$ atom finds itself in an already existing FM order between the NNN Mn--Mn. As the magnetic interactions in Heusler alloys are mediated via RKKY-type exchange interactions, the Mn--Mn interactions are predominantly influenced by the distance between them. This fact has been demonstrated using first principles calculations\cite{kubler}, as well as by using Monte Carlo simulations\cite{buchelnikov}, in Ni$_2$Mn$_{1+x}$Z$_{1 - x}$. Moreover, EXAFS and XMCD studies \cite{pab-epl, pab-prb} on Ni$_2$Mn$_{1.4}$In$_{0.6}$ also demonstrate the significance of bond-distance in dictating its magnetic interactions. Hence the new NN Mn--Mn$'$ interactions in Ga$_2$MnCo are AFM in nature. As the temperature is lowered, the lattice begins to contract and the insofar neglected NN interactions begin to gain relevance, setting a stage for competition between magnetic interactions. However, the number of new NN AFM pairs is much less (in the present case $\sim$ 10\% site disorder) compared to the FM pairs. Overall, a ferromagnetic order gets established in large spatial regions of the sample, with disorder induced AFM linkages between these spatial regions. The long range FM spin symmetry throughout the bulk sample is thus disturbed, giving rise to a cluster glass situation at low temperature.

\section{\label{sec:level7}Conclusions}
We studied the structural, magnetic and electrical transport properties of Ga$_2$MnCo Heusler alloy. Insights gained from neutron diffraction and ac susceptibility measurements help determine the magnetic ground state of this system. All the evidences suggest a cluster glass phase at low temperature and a ferromagnetic state at high temperature. The competing interactions between Mn atoms which at higher temperature administers a ferromagnetic type ordering, forms large clusters of FM interjected by tiny AFM interactions at low temperatures. The corresponding unusual distortion in the lattice, without change in symmetry is seen at the cluster-glass transition. The site-occupancies of magnetic atoms in the crystal structure play the critical role in controlling its low temperature magnetic ground state.

\section*{Acknowledgment}
TS and PAB gratefully acknowledge support from Department of Science and Technology (DST), New Delhi under project no: SR/S2/CMP-0109/2012. We thank Mr. Devendra Buddhikot (TIFR) for help in the measurement of transport and magnetic properties, and Mr. Ripandeep Singh (BARC) for help with neutron measurement. PAB also thanks the DST, India for the financial support and Saha Institute of Nuclear Physics, India for facilitating the experiments at the Indian Beamline, Photon Factory, KEK, Japan.

\bibliography{ref}

\end{document}